\documentclass[letterpaper, conference]{IEEEtran} 

\usepackage{cite}
\usepackage{graphicx}
\usepackage{listings}
\usepackage[T1]{fontenc}

\usepackage{hyperref}

\usepackage[table]{xcolor}
\usepackage[normalem]{ulem}
\usepackage{soul}
\usepackage{multirow}

\DeclareFontFamily{OT1}{phv}{}
\DeclareFontShape{OT1}{phv}{m}{n}{<-> s * [0.9] phvr7t}{}
\DeclareFontShape{OT1}{phv}{m}{it}{<-> s * [0.9] phvro7t}{}

\usepackage{color}
\usepackage{soul}
\definecolor{darkgreen}{rgb}{0,0.5,0}
\definecolor{purple}{rgb}{1,0,1}
\newcommand{\kibitz}[2]{\ifnum\Comments=1\textcolor{#1}{#2}\fi}

\newcount\Comments  
\begin{document}

\title{Organised Firestorm as strategy for business cyber-attacks}

\author{\IEEEauthorblockN{Andrea Russo}
  \IEEEauthorblockA{Department of Physics and Astronomy, University of Catania, Italy\\
    Email: andrea.russo@phd.unict.it}}
\maketitle

\begin{abstract}
Having a good reputation is paramount for most organisations and companies. In fact, having an optimal corporate image allows them to have better transaction relationships with various customers and partners. However, such reputation is hard to build and easy to destroy for all kind of business commercial activities (B2C, B2B, B2B2C, B2G). A misunderstanding during the communication process to the customers, or just a bad communication strategy, can lead to a disaster for the entire company. This is emphasised by the reaction of millions of people on social networks, which can be very detrimental for the corporate image if they react negatively to a certain event. This is called a firestorm.

In this paper, I propose a well-organised strategy for firestorm attacks on organisations, also showing how an adversary can leverage them to obtain private information on the attacked firm. Standard business security procedures are not designed to operate against multi-domain attacks; therefore, I will show how it is possible to bypass the classic and advised security procedures by operating different kinds of attack. I also propose a different firestorm attack, targeting a specific business company network in an efficient way. Finally, I present defensive procedures to reduce the negative effect of firestorms on a company.
\end{abstract}




\section{Introduction}
Before the advent of social medias, brand crises were largely caused by journalists’ contributions. 
Nowadays, a firestorm is a cluster of consumers' digital word of mouth that highlights some communication error, or some terrible mistake made by a company~\cite{hansen2018brand}. The Cambridge dictionary\footnote{\url{https://dictionary.cambridge.org}} defines the firestorm as \textit{``a sudden, and sometimes violent reaction''} and the shitstorm as \textit{``a wildly chaotic and unmanageable situation, controversy, or sequence of events''}. In this paper, I will use both these terms interchangeably.

During the last years, many firestorms took place on the Internet~\cite{pfeffer2014understanding,MCdonald,Astrazeneca}, mainly due to the increase of the number of users on social networks. In some cases, firestorms have been formally studied to better understand this phenomenon~\cite{nuortimo2020establishing, pfeffer2014understanding, hansen2018brand}. In 2007, several researchers debated over firestorms, and one of the main outcomes is that \textit{``a natural science model of the research process is suitable for studying the social world but a central issue remaining of whether the social world can, and should be, studied according to the same principles, procedures, and philosophy as the natural sciences''} ~\cite{BrymanBell2007}. This is relevant because today I are actually able to study and evaluate social dynamics by using the massive amount of data coming from the digital world, with particular emphasis on social networks~\cite{Rinaldi2017RealtimeMA}.

Firestorms are not made of a single event with a standard behaviour, instead they are caused by non-linear dynamics leading to complex behaviours. Due to this, companies must have appropriate procedures to respond to various crisis situations. Lehtonen's theory \cite{lehtonen1999kriisiviestinta} shows that a firestorm develops in five stages: \textit{(1) latent stage, where weak signals of the upcoming crisis are received; (2) triggering event, where the subject becomes the target of news and social media attention; (3) the subject is in the top-news and the media attention spikes; (4) the media attention calms down to the level of general philosophical and ethical discussion; and (5) there are only minor media hits and attention is guided to other issues} \cite{nuortimo2020establishing}.

As firestorms begin when there is a service failure, a social failure or when a company fails to communicate properly~\cite{hansen2018brand}, this kind of errors can be reduced by following appropriate procedures. However, most of the existing quality and security procedures, such as the ones suggested by ISO 9001:2015~\cite{ISO9001} and ISO/IEC 27002:2022~\cite{ISO27002} are not adequate for a multi-domain cyber and social attack. Because, regard to the 27002:2022, social attacks are outside the scope, while, 9001:2015, even if it focuses on better business process quality, thus, less firestorm risk from the public, it does not mitigate the firestorm from an attacker.




Hence, in this paper I theorise that it is possible for an attacker to intentionally cause a firestorm attack to undermine the reputation of a company, with the side-effect of advantaging the competitors. I argue that self-organised Firestorm attacks require a high number of bots that are already active on social medias: in this case, bots start the firestorm on the target company, spreading fake news (or magnifying a certain event, e.g., a mistake made by the company in the past) that will cause a high volume of real people to react negatively and continue the social attack, unknowingly on behalf of the adversary.

Additionally, I argue that Open Source Intelligence (OSINT) could allow an adversary to identify weak spots in the organization, namely people who most likely cannot react properly or defend themselves from the firestorm, hence not being able to timely mitigate its impact. Many workers have a LinkedIn, Facebook, or Twitter account: moving the firestorm on the social media accounts of people who work for the target company can lead to an extremely stressful situation for workers. This could be even worse for people who do not often deal with public relations, and could cause confusion, panic and distress. In fact, when a firestorm arises, even people who work on communication processes and managers can panic, and the fear of losing customers and partners can be very detrimental for any company.

When people working in the target firm are in this altered status, I argue it is possible to elaborate a social engineering strategy to capture protected information: in this case, not only firestorms serve the purpose to undermine the corporate image, but they are also used as a diversion for a social engineering attack. In fact, while most important organisations adhere to best-practices listed in security standards like ISO/IEC 27002:2022 \cite{ISO27002}, during a social attack like firestorms, some best-practices and procedures may be distorted or bypassed, both intentionally or by mistake, due to the pressure applied to people who are in charge of complying to such procedures \cite{halkos2010effect}.

\textbf{Contributions.} The paper makes these contributions:
\begin{enumerate}
    \item I explain how to make an automated and organized firestorm attack, with only a few manual operations such as the choice of a topic and of a hashtag;
    \item I introduce a taxonomy of possible actions that the attacker could perform while doing the firestorm;
    \item I illustrate how the author of a firestorm can evade detection for their attack by targeting single workers instead of the company profiles, while increasing the damage done to the firm.
    \item I show possible long and short term procedures that a company can implement to mitigate the effect of firestorms attacks.
\end{enumerate}

\section{Cyber-Attack Planing Prelude} \label{classAnalysis}
In this section, I illustrate a novel strategy to artificially cause a firestorm, leveraging a botnet to start agitating real people against a target company.   
Due to the large number of posts that bots can create within seconds, they can be used to amplify any idea on social networks, influencing political affairs~\cite{carrerandbechis2020} and business company value~\cite{barcellona_2021}.
For example, due to a cyber-attack on a Twitter newspaper profile, such newspaper shared a fake news about President Obama being injured by a bomb in the White House, causing a flash-crash in Wall Street and the stop all of economic transactions for some minutes. This led to a loss of about 121 billion dollars for S\&P 500 and its related companies~\cite{Flashcrash}.

I structure the attack plan in six stages:

\begin{enumerate}
    \item \textbf{Finding an event/topic to build the firestorm attack on.} This can be a past event or an error that the firm has committed in the past, which will be used as a basis for the upcoming attack. I define this event as the \textit{target topic}.
    
    \item \textbf{Using bots to create or amplify the latent state.} By leveraging a botnet, an adversary can create a high number of posts on social media, allowing the target topic to reach more people and giving them the opportunity to react negatively. This can eventually lead to a state where real people start to autonomously talk about the subject and begin to spread information about the target topic on their own. To facilitate this, the attacker can reuse an old trending hashtag or create a new one: the hashtag is the keyword to incite social action due to the information symbolised by the word itself. 
    
    
    \item \textbf{Letting the topic spread among people.} The ideal situation for the attacker is that real people begin posting about the target topic, after learning about it from the botnet's posts. This will bring more attention to the topic, possibly making it a trending one. For example, Twitter allows users to check what topics and hashtags are currently popular. 
    If this happens, there will be moment in which there are enough people posting about the target topic, so that the firestorm can sustain itself for days, without any other post coming from the attacker's botnet. I call this moment the \textit{fire point}.\footnote{In chemistry, the \textit{fire point} is the lowest temperature at which a certain fuel will continue to burn for a minimum of five seconds, when ignited.} Instead, if real people did not react negatively to the topic, or the topic did not reach enough people to allow the firestorm to reach the fire point, the discussion on the topic will slow down and will eventually end. In this case, I say that the firestorm is \textit{extinguished}. However, the attacker can change the target topic and restart from Stage 1.
    
    \item \textbf{Identifying human targets.} Managers (e.g., Chief Technical Officers, Chief Executive Officers) are the decision makers of a company. The attacker might want to keep a list of these people in order to use these names when the attack will move over from the company's social network profiles to the employees' ones. Identifying the people who are most proud to work for the attacked company can also be helpful in exerting more pressure on the company (since they have more to do with the value of the company). 
    
    \item \textbf{Focusing on workers.} During the peak activity of the firestorm, those same bots that built the latent state will move their focus on the public social media profiles owned by employees of the attacked firm. These profiles were identified in the previous step of the attack. This may cause the attention of the firestorm to shift towards the employees, also causing them to experience discomfort. Because the brand is usually at the center of the firestorm, focusing people will have a stronger impact on them, and it can disrupt internal processes.

    \item \textbf{Performing the cyber attack.} Because people will put less attention in following internal procedures, many safety best-practices adopted by the company may not be followed properly, or may even be ignored. The attacker can exploit this behaviour to their own advantage.
\end{enumerate}

In order to shift the focus from the company to the worker, it is necessary to optimise the timescale and timing of the transition, as it is not linear for people to attack the worker, but it can happen more easily if the negative event is of high negative impact and value.
Shifting the attack on employees has another side-effect, which is beneficial to the attacker: the organisations that are responsible for the public cyber security in every country cannot see the Firestorm attack on the company page, because the Firestorm is focused on workers only
Such organisations will hardly be able to detect all comments and posts focused on workers, allowing the attacker to create a smoky form of the attack, which can bypasses conventional security measures, procedures and strategies. Since they have to focus primarily on the company under attack, therefore, possibly not give so much attention to analysing every single interaction against all the operators of the attacked company.
\\


\section{Business Social mood-disease and network strategy} \label{testGeneration}

The Cambridge Analytica case highlighted the role and the importance of social media for the majority of the population and organisations. A document produced by the American Ministry of Justice, to examine the possible foreign influence on US, showed how there actually exist organisations (such as the IRA - Internet Research Agency)~\cite{intelligencesenate} that aim to influence individuals, public and private organisations~\cite{ReportUSA}.

A great part of what is needed to successfully influence people lies to understand the initial conditions of the system, i.e. in the correct profiling of such people through data obtained on social networks.  People who are more sensitive to certain issues, and those key people who can influence the most the community where they live and work are the main focused people for a social attack, because they have a central role (hubs) in the network.  

Profiling consists in obtaining (through a process of data collection and subsequent processing) an absolute or almost absolute understanding of a group of individuals or a single person, comprehending their habits and preferences \cite{Profilazionesicurezza}. The information obtained concerns political, musical and social interests, including the identification of their network of friends, colleagues, and much more. 
This information allow a much easier conveying of any content, as it is possible to understand who is most susceptible and interested on various topics, affecting their weaknesses, fears and interests. Furthermore, it is possible to infer who could possibly propagate a certain content through their network, exponentially increasing the chance of success if the subject in question is a person with an important or main role.

Cambridge Analytica used the OCEAN model, related to personality traits, to understand preferences of many people in the US during the national election on 2016 \cite{intelligencesenate}. The OCEAN model allows to send specific messages and contents to people who are sensible to a certain topic. This method is very different from the classic and standard mass communication, because it is possible to send the right content to the right person. Unfortunately, the CA scandal was defined as classic political influence, the old-fashioned way, thus including prostitution, favouritism, etc.  In reality, the scandal found ``a new type of weapon'' as Brittany Kaiser (former CA business development director) said during her question time (on Commons culture committee in 2018)  to describe the work done from CA, but also to categorize AI as a real soft-power weapon~\cite{Profilazionesicurezza}.

However, understanding hot topics for workers is not enough -- in order to modify their mood and obtain a good social attack, a subject topic needs to be found as well. 
On social networks, during firestorms 
, people are usually triggered by three kinds of errors \cite{hansen2018brand}:

\begin{enumerate}
    \item Social failure
    \item Communication failure
    \item Product or service failure
\end{enumerate}

Although they may seem similar, different types of events can lead to different types of dynamics and reactions. 
In the case of product or service failures, for example, performance-related crises raise doubts about the brand's ability to deliver basic functional performance \cite{dawar2000impact}. Another research has also identified not only short-term effects to a brand after a firestorm, but also measured long-term ones, at least two years after the latest firestorm \cite{hansen2018brand}. 

I hereby give an example for each of the aforementioned triggering factors. 

\begin{enumerate}
    \item \textbf{Social failure.} The firm might be an accomplice of some accident or crime, like Nike with children shoes~\cite{nike, Brent} or the ING-DiBa case in 2012~\cite{pfeffer2014understanding}.
    \item \textbf{Communication failure.} The firm might fail to communicate properly, for example making negative comments regarding a certain community or movement~\cite{MCdonald}.
    \item \textbf{Product or service failure.} The firm might distribute a product that harms consumers, for example a vaccine that can kill people~\cite{Astrazeneca}.
\end{enumerate}

These failures and the firestorm stemming from them might cause affected employees to experience discomfort and panic, because coworkers, friends and other people in their network might see affected employees as the root-cause of the Firestorm. 

The social-cyber attack also provokes unlikely passive consequences for companies: 
\begin{enumerate}
    \item The value of the company on the financial market could rapidly decrease; \cite{Flashcrash} 
    \item People who worked in the company during the firestorm might be subject to discrimination in future, especially if the firestorm was caused by a (supposedly) unacceptable mistake that could have been avoided~\cite{scottishleft,VolkswagenRallied}. 
    \item As the people, also the offended brand could carry a long-term stigma that would motivate other companies to make job offers to the personnel of the attacked firm. This could put it on an even greater disadvantage, as workers would be incentivized to leave the attacked company and accept the new offer. 
\end{enumerate}

The network, as well as the importance and scope of the news, can thoughtfully influence the reaction and dynamics of the company. 
The network, as well as the importance and scope of the news, can thoughtfully influence the reaction and dynamics of the company. For example, when a company's workers receive an high importance news, they may behave helplessly in relation to the importance of the news; feeling relieved of responsibility, since the event is bigger than their actions, they tend to pass much of the responsibility on to the company's managers. 

Indeed, in times of disorder or chaos, Entropy increases with decreasing order, and emergency increases with increasing order: this happens because people within the organisation understood the emergency, and the organisation improve them-self to respond to it~\cite{Entropyemergence}.

When many workers in the company are panicking, the organisation's CCO (Chief Communication Officer) will elaborate and react to Firestorm on company pages, however, this cannot stop the social attack on the individual profiles of the employees. Hence, even people who are in charge of running communication processes and managers can panic, as the more is the duration of the firestorm, the higher is the chance of losing clients and reputation. This is a terrible situation for any company, especially after many years of work.
However, managers are considered "critical workers" on the organisation chart, hence, they cannot be influenced by social manipulations and social diseases, because of the responsibilities they have in the company. 
While during the last century such organization charts had the form of a pyramid, usually with the CEO on the top, nowadays the AGILE model allows companies to organise their personnel in different ways within their organization charts. However, the legal and personal responsibility for every error or critical issue will be always be of the top manager of that area -- for example, the CISO (Chief Information Security Officer) is usually responsible for the cyber security.
A network side strategy can hard-influence workers close to managers and directors, contaminating directly the mood of the team, including the manager. In a more specific way, the attacker the hub from the company network, defusing also other workers from the company.
Once the social-disease is already widespread on the company, and many people are stressed about the firestorm, the cyber attack can begin.

   
\section{Assessing the Attack Surface}
In this section, I introduce the possible actions that the adversary (or the real people that contribute to firestorm) can perform to further disrupt the target company's business processes, to sink its corporate image, or to get classified information. To do so, I introduce a novel classification of these actions and analyze their impact on the fundamental properties of information security, that is, Confidentiality, Integrity and Availability~\cite{samonas2014cia}.

I show these actions can be divided in three categories:

\begin{enumerate}
    \item \textbf{Controlling Large Scale Entities}, that is, thousands or even millions of different actors performing several concurrent actions against a firm. These actors can act both remotely and physically, and can be both robots and humans.
    \item \textbf{Leveraging Internal People}, namely, exploiting mistakes performed by employees (e.g., because they are stressed due to the firestorm), or having an insider threat who can extract classified information.
    \item \textbf{Asking for Ransoms}, that is, the adversary may want to ask for a payment to stop the firestorm. This would cause the bots to be shutdown, or even to defend the company on social medias.
\end{enumerate}

I hereby analyse the different actions within each category and their impact. This analysis is summarised in Table~\ref{tab:attack-surface-assessment}.

\subsection{Controlling Large Scale Entities}

\paragraph{Denial of Service (DoS) Attacks} The adversary might want to harm the firm's reputation by negating the availability of the services it offers. To this avail, the attacker can leverage botnets to send a very high number of requests per second to the target service, overwhelming the server and resulting in the service going down. If possible, the attacker could even reuse the botnet used to create the latent state, and rearm it with a DoS script. Alternatively, if the adversary is not a single entity but a large group of organised people, a DoS attack can be performed with simple scripts, without leveraging any botnet, as the large number of adversaries could be able to generate the traffic required to overload the server. In this case, however, the adversaries would have to carefully time their attack, and they might want to hide their location, for example by using a VPN. Finally, the adversary could encourage real people to overload the target firm's servers, as they could co-ordinate the attack by using the bot profiles used for the hashtag propaganda.

\paragraph{Physical Actions} Business processes can be also interrupted or slowed by legal, yet harmful, physical actions. One example is a demonstration around the firm's premises: employees might not get to their workplace in time because people manifesting outside the building are blocking or slowing access to the premises, or they are creating more traffic than usual on the way to the building. Another example is people calling the organisation's call centers with the only goal of protesting.  

\subsection{Leveraging Internal People}

\paragraph{Human Error} Even though it is widely known that human error is one of the most prominent causes of security incidents~\cite{HumanError1993, HumanFactor2021}, most companies still do not adequately invest in training for their personnel, resulting in data breaches or other security related events~\cite{langlois20202020}. This means that, if the attacker wants to obtain an initial foothold on the target organization's systems, they might be able to do so without needing a firestorm attack, depending on the employees' ability of recognizing phishing emails or scam websites. However, workers who are experiencing firestorm, be it on the company they are working with or on their own profile, will be more inclined to break internal policies, hence committing mistakes, due to the perceived crisis~\cite{bakos2019human}.

\paragraph{Offering Help} During the firestorm's peak activity, the adversary itself contacts the attacked firm, pretending to be a professional (e.g, a consultant) who can help in mitigating the effects of the firestorm, for example as a Social Media Manager who has dealt with Firestorms before. This can happen via emails, social networks or through the corporate's website, for example if the firm has some job openings and the adversary pretends to be a candidate. For smaller enterprises, the adversary may even show up in person to the attacked company's premises. If the attacker manages to get hired, they might get access to classified information. I argue the attacker does not want to tamper with documents or attack the firm's infrastructure while being an employee themselves.

\paragraph{Insider Threats} Instead of joining the firm themselves, the adversary might establish a contact with employees who are still in the attacked company but are not showing support on social media, or even manifested dissatisfaction towards the company. The attacker might want to try to persuade them in sharing confidential information, making them insider threats~\cite{insiderthreatsincs} -- if they have success, not only they acquire classified information, but if the stolen content is also compromising for the firm, it could be published online to damage the firm's reputation even more.

\subsection{Asking for Ransoms}

\paragraph{Extortion to Stop the Attack} The adversary contacts the attacked firm and proves the botnet that is performing the firestorm is in their control. They then ask for an arbitrary amount of money in Bitcoins to shutdown the bots, stopping a (hopefully) substantial part of the attack. In fact, if the firestorm already managed to incite many people in joining the social attack, the shutdown of the botnet might not stop or slow down the firestorm. If the adversary plans to attack multiple firms with their firestorms, they to avoid situations like this, because the odds of a victim paying a ransom is proportional to the reliability of the attacker in stopping the attack once they receive the money. In other words, the attacker must be considered ``trusted'' in stopping the attack if the ransom is paid, so victims are more incentivized to pay~\cite{cartwright2019pay}.

\paragraph{Defence as a Service} The adversary contacts the attacked firm, but instead of showing they are in charge of running the attack and asking money to stop it, they try to sell a fire(storm)fighter service to the victim, supposedly consisting on bots defending the reputation of the firm: this is basically a reversed firestorm, in which those same bots that built the latent state now defend the company: to avoid drawing excessive attention, the attacker might slowly change the proportion of attacking bots versus defending ones, until they are all defending the company.

\begin{table}[!ht]
    \caption{Social Attack Surface Assessment}
    \footnotesize
    \label{tab:attack-surface-assessment}
    \begin{center}
    \resizebox{\columnwidth}{!}{
    \begin{tabular}{ l l c c c c }
        \hline 
        \multirow{2}{*}{\textbf{Category}} & \multirow{2}{*}{\textbf{Action}} & \multicolumn{4}{c}{\textbf{Impacts}} \\ \cline{3-6}
        & & \textbf{Confid.} & \textbf{Integ.} & \textbf{Avail.} & \textbf{Rep.} \\
        \hline
        \multirow{2}{*}{Large Scale} & DoS Attack & No & No & Yes & Yes \\
        & Phys. Actions & No & No & Yes & Yes \\
        \hline
        \multirow{3}{*}{Internal People} & Human Error & Yes & Yes & Yes & Yes \\
        & Help Offer & Yes & No & No & No \\
        & Insider Threat & Yes & No & No & Yes \\
        \hline
        \multirow{2}{*}{Ransoms} & Extortion & No & No & No & No \\
        & Defence Service & No & No & No & No \\
        \hline 
    \end{tabular}
    }
    \end{center}
    
    \footnotesize{\textbf{Confid.:} The action can affect the Confidentiality property. | \textbf{Integ.:} The action can affect the Integrity property. | \textbf{Avail.:} The action can affect the Availability property. | \textbf{Rep.:} The action can negatively affect the reputation of the company.}
    
\end{table}

\section{Case Study: CD PROJEKT RED}
\begin{figure}
	\centering
		\includegraphics[width=\linewidth]{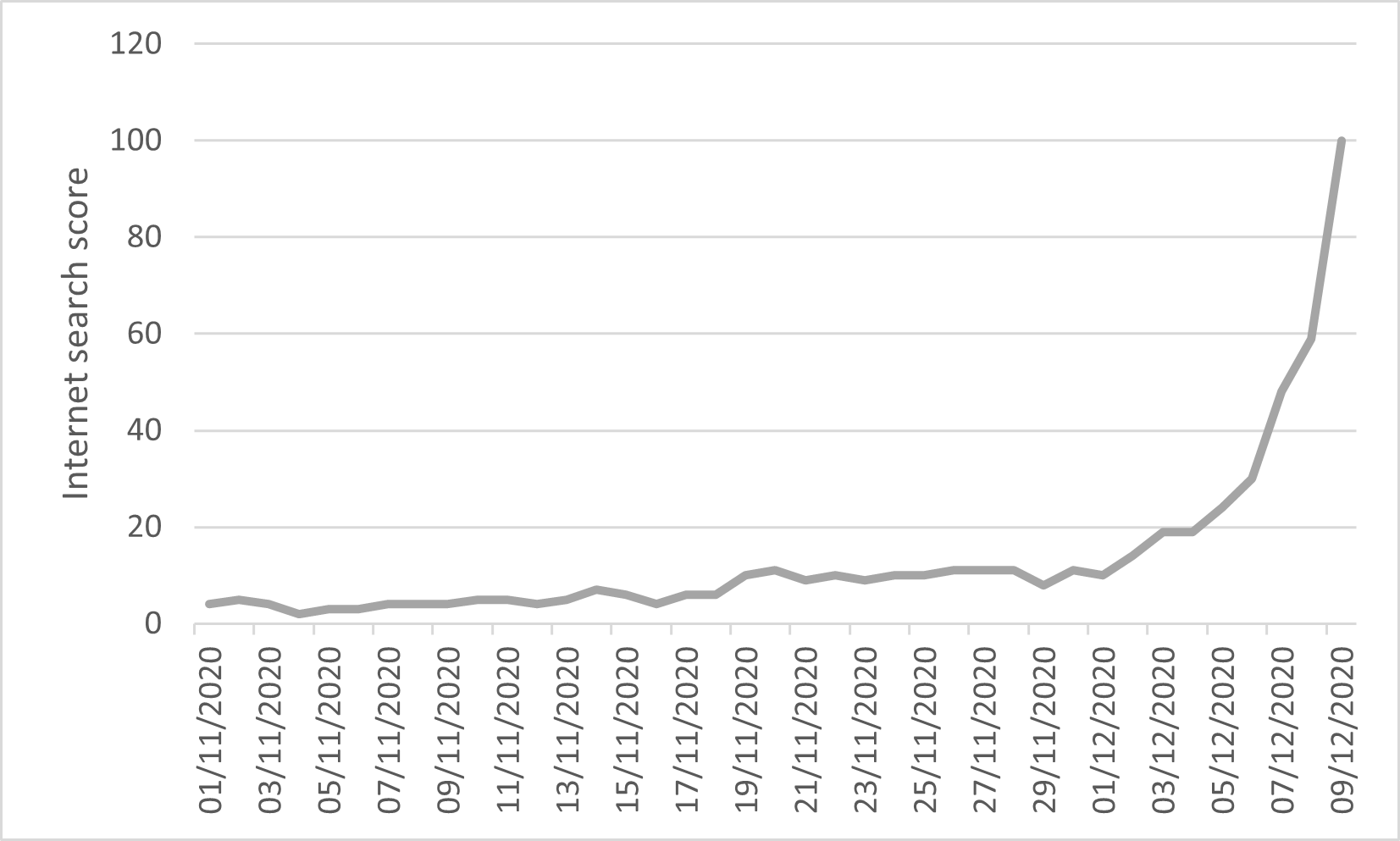}
	\caption{Interest Score showing social hype for the release of Cyberpunk 2077}
	\label{FIG:1}
\end{figure}

\begin{figure}
	\centering
		\includegraphics[width=\linewidth]{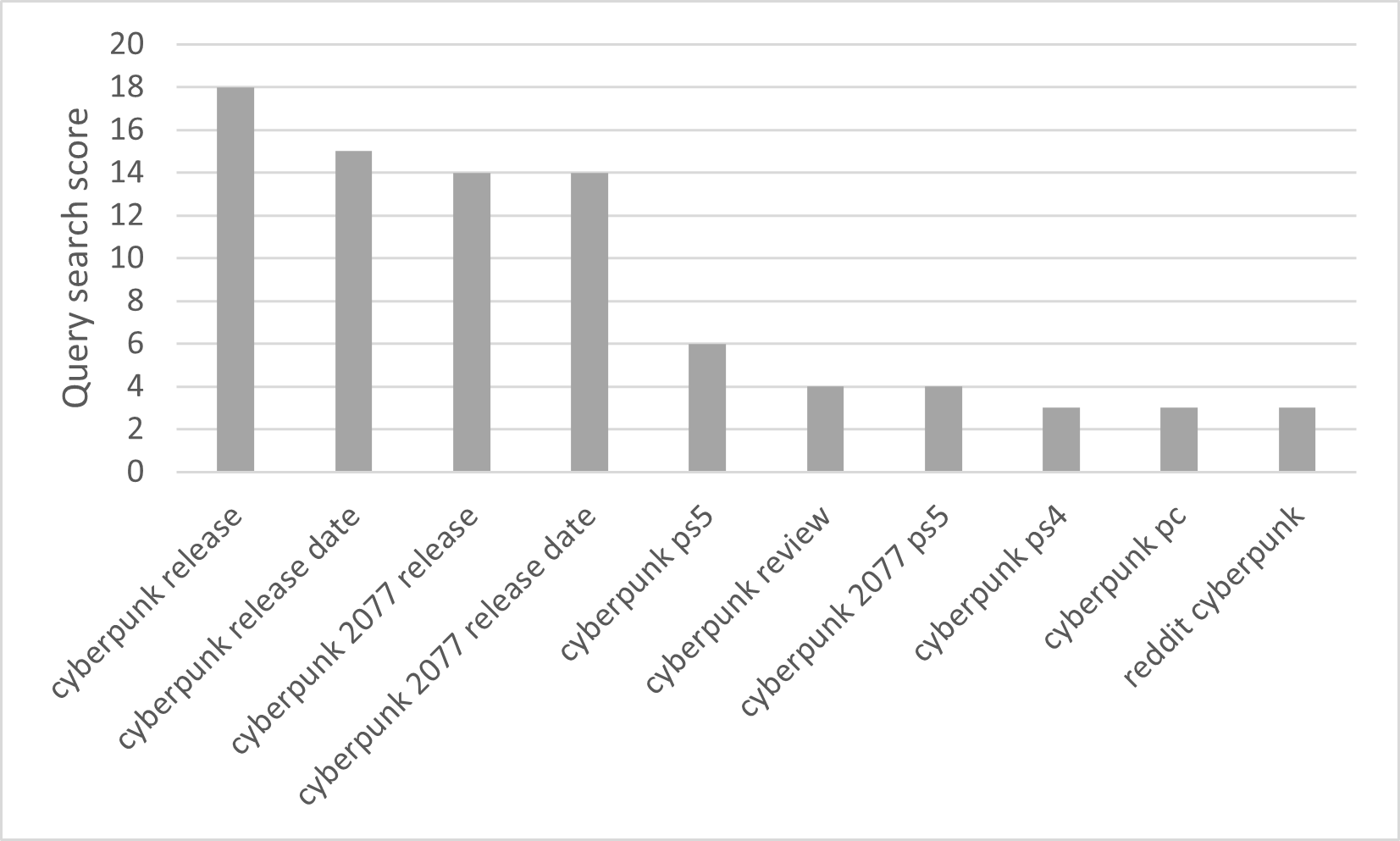}
	\caption{Queries showing social hype for the release of Cyberpunk 2077}
	\label{FIG:2}
\end{figure}

On December 10, 2020, CD PROJEKT RED released a long awaited game called Cyberpunk 2077. This game was very popular even before its release and it generated continuous social hype from the video game community throughout its development, also winning the ``Best Game Awaited'' from Golden Joystick Awards for two consecutive years.\cite{Wikipedia_CD}
As shown on Figure~\ref{FIG:1} and Figure~\ref{FIG:2}, hype for the game substantially increased during the 10 days before the release of the game, reaching its apex on December 10, when the hashtag \texttt{\#Cyberpunk2077} was tweeted 193,900 times on Twitter, from users of 53 different nationalities. During this time span, many other hashtags regarding the game were very popular, for example \texttt{\#Cyberpunk2077Hype} was retweeted 10,000 times \cite{Getdaytrends}.

However, a few days after the release , the Cyberpunk 2077 topic arise again, this time associated with queries related to patches and refunds. In fact, the game was released too early and many bugs were present: due to this, several people had asked a refund to CD PROJEKT RED, often also writing a bad review for the game on online stores.
This created a "information-disease" 
within the company, just like the one described in Section~\ref{testGeneration}: in this case, CD PROJEKT RED's employees became stressed and felt pressure related to the quality of Cyberpunk 2077, in which they had invested more than two years of hard work. \cite{Wikipedia_CD}

In early February 2021, only 60 days after the game's release, CD PROJECT RED was hit by a ransomware attack and attackers were able to extract the source code of several games, including administrative files~\cite{CDproject}. The attackers then threatened the company of leaking or selling the stolen code and files, unless the firm paid a large amount of money to the cyber-criminals. In the end, CD PROJECT RED refused to negotiate with the attackers, stating on a press release that they would ``not give in to demands or negotiate with the actor'', also confirming that no personal information was obtained in the attack and that they were working with law enforcement to track down the attackers~\cite{CDproject2,CDproject3}. Later on, security analysts found the stolen source code while being auctioned on the dark web for a minimum price of 1 million USD. \cite{Arstechnica} The auction was closed after the attackers stated they had received an offer that satisfied them \cite{Arstechnica} Within a week of these auctions, the code was shared online via social media, and CD PROJECT RED began using DMCA take down notices to remove posts containing their code \cite{Vice}.

The social hype that CD PROJEKT RED generated for Cyberpunk 2077, was used by hackers to threaten the company in order to extorting money, but also, had a side effect, i.e. damaging the company's reputation, that can bring to undermine the sales of other long awaited games. 

In Table~\ref{table:Vader} I show the results of the sentiment analysis, obtained from tweets and comments for the hashtag \texttt{\#CDprojectRED}. Data collected from Twitter respects the timeline of Cyberpunk 2077's release and its development; 
data shown in the table can be organised in three categories: before release (October and November), during release (December and January) and after the release of Cyberpunk 2077 (February).

It is possible to observe that in October and November the sentiment remained neutral-positive with a few oscillations. In December, when the game was released, I can observe a small increase in the negative sentiment due to the high number of bugs present in the game, however, this increment is quite negligible. In January, when a greater number of players were playing the game, the negative sentiment became stronger than the positive one, causing not only a negative compound (-0.111), but also a neutral-negative sentiment for the game and for the developers. Finally, on February the sentiment returned neutral overall, however, the presence of negative sentiment is still stronger compered to the one in October and November.

\begin{figure}
	\centering
		\includegraphics[width=\linewidth]{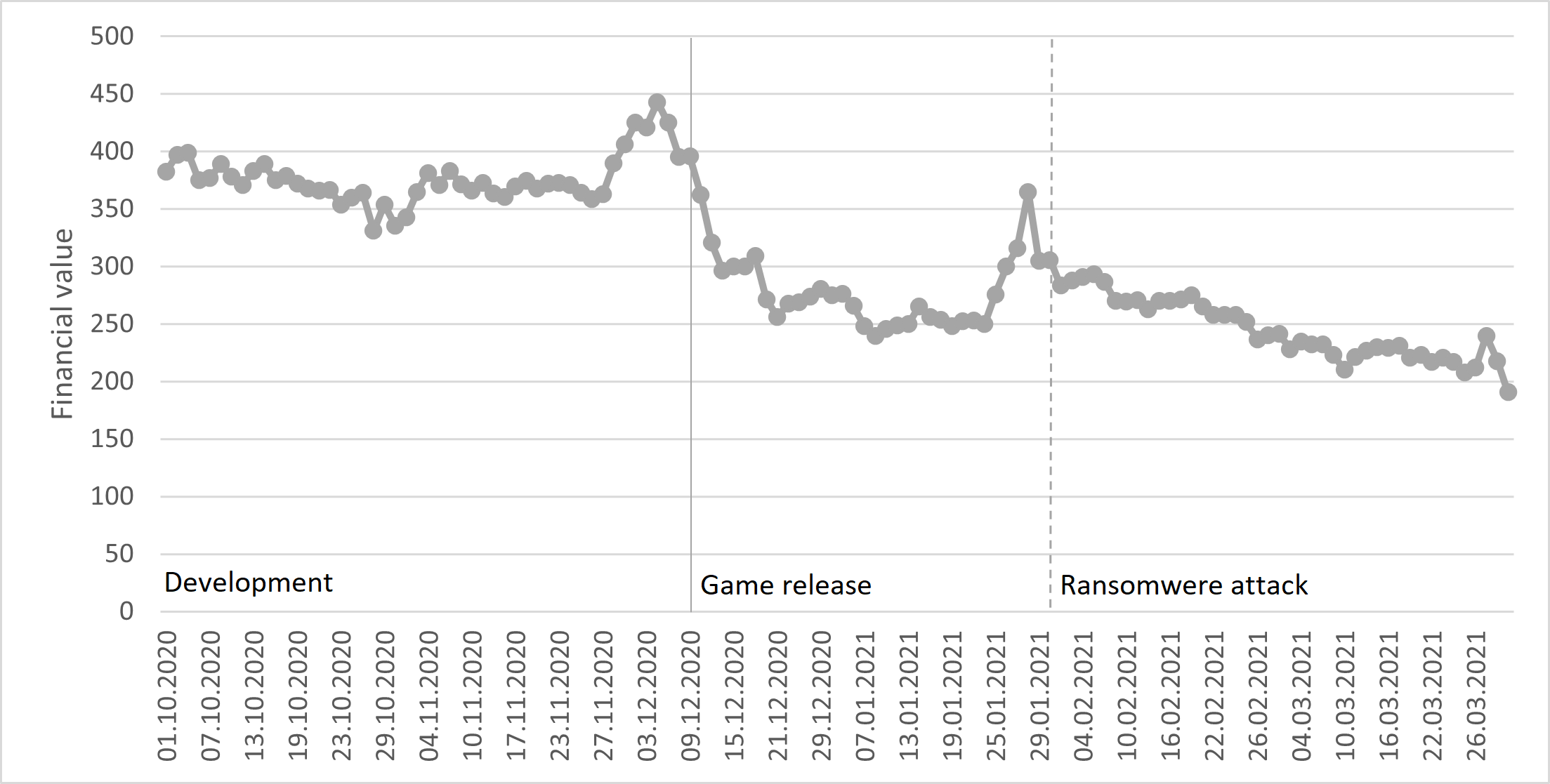}
	\caption{Financial value of CD PROJEKT RED and critical events}
	\label{FIG:3}
\end{figure}

These data show how much pressure the CD PROJEKT RED company had to experience during the release of the game. Additionally, in Figure~\ref{FIG:3}, I show the financial value of the company during the whole game release timeline, also marking the two critical events that occurred: the yellow line indicates the release of the game, while the red line indicates the ransomware attack. I can see that, after the release of the game, the financial value of the company suffered a sudden drop, that was likely conditioned by customers losing trust in the company due to the presence of many bugs in the game, bad reviews and critics. I can see that the company regains more than half the value lost during the next two months, however, the ransomware attack causes another drop in the financial value of the company due to customers losing trust in the company again, this time from a security perspective.

\begin{table}[ht] 
\caption{Vader Sentiment on \#Cyberpunk2077 from Twitter} 
\centering       
\begin{tabular}{l c c c c} 
\hline
Months&Negative&Neutral&Positive&Compound\\ [0.5ex]  
\hline                    
October & 0,085 &	0,757 &	0,150 & 0,163\\     
November & 0,079 & 0,766 &	0,149 &	0,163 \\ 
December & 0,087 & 0,750 & 0,161 & 0,153  \\ 
January & 0,143 & 0,758 & 0,093 & -0,111 \\ 
February & 0,104 & 0,745 & 0,145 & 0,120 \\         [1ex]
\hline      
\end{tabular} 
\label{table:Vader}   
\end{table}

\section{Business defence strategy} \label{experiments}
To avoid dangerous events for companies, human factor is a crucial element \cite{formicheumano}, however it is also possible to create specific defence strategies. Failures introduced in Section~\ref{testGeneration}, i.e. social failures, communication failures and product or service failures can be analysed to prevent incidents. To the most of us, the news that a particular piece of information (e.g. a meme, a hashtag) went “viral”, reaching millions of nodes in a short period of time may seem purely random and hence unpredictable, but Kolli et al.~\cite{KolliCascade} discovered that, at least 20\% of the times, the cascade volume changes in a manner that appears to be random, and in the remaining 80\% it is possible to predict the cascade’s future volume. Hence, it is possible to create short-term strategies to detect firestorm attacks while they are still in the early stages, i.e. while the latent state is being built. 
However, it is also possible to create long-term defence strategies with a proactive governance. A possible proactive strategy for the long-term could be as follows:

\begin{enumerate} 
    \item Organise internal company procedures to help employees protect themselves against various attacks on social media (like Linkedin); 
    \item Organise procedures outside the company, such as contacting allied/partner companies for help with the various attacks on social media;  
    \item Create in advance supporting bots that will defend the company automatically; 
    \item Create an international database of accounts that have made firestorm. The database, accessible to all organisations, both public and private, will help to understand whether the type of firestorm taking place is real or artificially created.  \cite{cyberthreatintelligence}
\end{enumerate} 


These three possible actions can be highlighted by the mass media, which will publicly show that the firestorm is being fought because other people or organisations began defending the attacked company. Hence, these actions allow the firestorms to calm down, and eventually to be extinguished, faster than simply doing nothing.\cite{hansen2018brand} 
If a company has done something enormously wrong in the past, it is possible that every time the same company does something wrong, there is a chance that another firestorm can restart, either for the recent event or also for the past one. In fact, the firestorm can come back with an interval of about 2 years \cite{hansen2018brand}. 

In case of social failures, there is also an additional side-effect that must be mitigated, that is, the firestorm naturally expands to the employees without the manipulation of the adversary. Example defence strategies against this side-effect could be implemented as follows:

\begin{enumerate}
  \item{Let people from outside and inside the company on social network, dialogue about that topic (such as the case of carnivores vs vegetarians at ING-DiBa \cite{pfeffer2014understanding}). This strategy can increases the number of followers;} 
   \item Blame an entity that is external to the company as a scapegoat, so the Firestorm can move from the company to the designed entity. Even if it is not very moral, it is something that usually works; 
   \item{Depending on the strength, length, and breadth of the attack, it is possible to make strategy about possible reaction for company. 
   }
  \begin{enumerate}
    \item Social failure: If the firestorm is linked to a partner company, or only a certain sector of the company is under attack, immediately distance yourself from them. 
    \item Communication failure: The goal here is to safeguard the company's reputation and authority. In this case, try to detach yourself immediately from the communication error, and continue with the company's reputations strategy, making it appear that it was just an accident on the road. Furthermore, apologising for the event never hurts.  
    \item Product or service failure: Instantly block the production of the affected product or the provision of the service. Organise a commission that can evaluate the quality of product/service. Even if it is complicated given the amount of partners, quality standards and corporate continuity, this action, if done in time, creates a good defensive shield at the communication level, as people can understand that the company itself has also understood the problem, limiting the damage;  
  \end{enumerate}
 \end{enumerate}
 
Timing is essential during Firestorms, first of all to understand whether the type of firestorm is real or artificial (you can tell by the date of creation of the accounts that do firestorm -- if the initial accounts were born recently, they are probably bots, hence artificial); secondly for improving the cyber defence and be prepared for a possible cyber attack; tertiary for the public reaction, because it means that the affected company has noticed the failure faster or as fast as other people (who are doing the firestorm on social networks) and will promptly react to the problem, reassuring customers that it will be solved. This will help in calming down or extinguishing the firestorm. For example, the carnivores vs vegetarians case at ING-DiBa was caused by a communication failure. The company had never had so much traffic on its Facebook page before, and they saw in this an opportunity to increase the number of their followers. In fact, after a few days had passed from the firestorm, and the attackers were still posting, newly-acquired followers jumped into the debate and started defending the company. \cite{pfeffer2014understanding}\\
\\
Obviously, depending on the type of firestorm,real or artificial, it is necessary for the company to adapt its strategies according to the type of attack (real or artificial). The prevention part, of course, works in both cases, but understanding who you are fighting against and the causes, helps to save the reputation of the company, and sometimes even the company itself. 



\section{Future Work} \label{related}
In one of the next jobs, I would like to implement different pressure dynamics, i.e., either implement rapid, massive, and incisive firestorms, or permanent, with few accounts firestorm. 
Depending on the firestorm, these types of dynamics can change the pressure on companies and workers in different ways, perhaps showing that for some companies it is better to have a permanent firestorm, or for others a rapid one.  
Another aspect I would like to draw attention to in future work is also how people are contacted in the company, i.e. with messages that are more likely to provoke an ethical reaction, for example, when people are contacted by bots and they point out to the worker the disaster he has made to his company. 
This case is very interesting, as it is possible, after 'moralising' the worker, to apply social engineering strategies to facilitate the cyber attack. 
On the other hand, on the side outside the company, i.e. not focused on employees, strategies can be used to increase the chance of a successful cyber attack, or extortion of information or money. For instance, during the firestorm, it is possible to contact the company under attack, and pose as the national cyber security agency, initiating strategies such as: 
\begin{enumerate}
    \item Passing themselves off as the national cyber security agency, they say that most are fake accounts and get information on their security; 
    \item Passing themselves off as the national cyber security agency, enter in their computer system.
    \item Passing themselves off as the national cyber security agency, saying they are carrying out a cyber attack to test their cyber defences, carry out a second attack immediately afterwards, exploiting the information from the first attack and passing on part of the defences, or, say they are not defending themselves against the first attack so as to obtain the desired data. 
\end{enumerate}
In any case, these kinds of interactions will be carried out by means of computer simulations, since for obvious ethical reasons it is impossible if not extremely difficult to apply these strategies.

\section{Conclusions} \label{conclusions}

In this paper, I have shown how some events related to cyber security are linked to certain social dynamics. When social dynamics are mixed and linked to cyber purposes, classic attack types (cyber or social attack) can no longer be defined, but social-cyber attacks, as the effectiveness of one also induces a probability of success of the other. \\
I introduce an novel model allowing researchers and companies to (1) understand when companies and organisations have fragile defence against a social-cyber attack, (2) illustrate how company and organisation can defence them self from firestorm, (3) proving that social-cyber attack must be defined as a possible high risk event as multi domain sector, and (4) showing a now model of cyber attack, with a multidisciplinary sociological approach to increase the potentiality of common cyber attack.
The data collected from CD project red's event case, shows how these types of attacks, although still little known, may become a norm in the future, as the company's assets are not only its human capital, or the production of goods and/or services, but also its own reputation. 

\section{Authors \& Paper Information}

\subsection{Data gathering}
I collect tweets related to the topics \#Cyberpunk2077 by using Tweepy and the Twitter archive API.
Both service use the permission from Twitter to obtain and gather data, but any downloaded topic need revisions and cleaning process to increase the quality of the research. For example, I found many copy-paste tweets (caused by spamming process, or fake-account/bot), and also several tweets had (during the Vader Sentiment Analysis) incomprehensible word for the Vader program, and I deleted it. 
For any topic I use the same methodology to obtained standard and quality data. 
In addition, to obtain the correct amount of tweet (define as the number of tweet) for each day/hour I use getdaytrends.com, a specific site where it is possible to monitoring every topic in real-time and also aged topic.
In total, our data count more then $\sim$5000 Tweet.
I obtain the Financial data of CD project RED from https://www.investing.com/equities/cdproject-historical-data site.

\subsection{Author Contributions}
Investigation and data resources, methodology, data cleaning and software, A.R.;
All authors have read and agreed to the published version of the manuscript.

\subsection{Funding}
The author(s) disclosed receipt of the following financial support for the research, authorship, and/ or publication of this article:
This project has received funding from the University of Catania. 

\subsection{Author biographies}
Andrea Russo is a PhD candidate in Complex Systems at the University of Catania. 
He is currently working at the Department of Physics and Astronomy. He collaborated with CNR Ibam, he also has worked purely on projects involving technology and society. \\
His main research field and interests are focused on the study and the development of Computational social method to explain social complexity, in particular field like Politics - Economics - Business and Defense-Security sector applications. \\
Orchid: 0000-0003-3816-0539\\
Corresponding author. Email: Andrea.russo@phd.unict.it\\

I would like to thank "Vereos" and "Andrea metal clone", who helped me in idealising and refining the paper. 


\nocite{*}
\bibliographystyle{abbrv}
\bibliography{refbib}





\end{document}